\documentclass[12pt,preprint]{aastex}

\shorttitle{Black Hole SASI}
\shortauthors{H. Nagakura and S. Yamada}

\begin{document}

\title{General Relativistic Hydrodynamic Simulations 
\\and Linear Analysis of the Standing Accretion Shock Instability around a Black Hole}

\author{Hiroki Nagakura \altaffilmark{1}and Shoichi Yamada\altaffilmark{1,2}}

\altaffiltext{1}{Science and Engineering, Waseda University, 3-4-1
Okubo, Shinjuku, Tokyo 169-8555, Japan}
\altaffiltext{2}{Advanced Research Institute for Science and Engineering, 
Waseda University, 3-4-1 Okubo, Shinjuku, Tokyo 169-8555, Japan}
\email{hiroki@heap.phys.waseda.ac.jp}

\begin{abstract}

 We study the stability of standing shock waves
 in advection-dominated accretion flows into a Schwarzschild black hole
 by 2D general relativistic hydrodynamic simulations as well as
 linear analysis in the equatorial plane.
 We demonstrate that the accretion shock is stable
 against axisymmetric perturbations
 but becomes unstable to non-axisymmetric perturbations.
 The results of dynamical simulations
 show good agreement with linear analysis
 on the stability, oscillation and growing time scales.
 The comparison of different wave-travel times
 with the growth time scales of the instability suggests that
 the instability is likely to be of the Papaloizou-Pringle type,
 induced by the repeated propagations of acoustic waves.
 However, the wavelengths of perturbations are
 too long to clearly define the reflection point.
 By analyzing the non-linear phase in the dynamical simulations,
 it is shown that quadratic mode couplings precede
 the non-linear saturation.
 It is also found that not only short-term random fluctuations
 by turbulent motions but also quasi periodic oscillations
 take place on longer time scales in the non-linear phase.
 We give some possible implications of the instability
 for quasi periodic oscillations (QPOs) and
 the central engine for gamma ray bursts (GRBs).

\end{abstract}


\section{Introduction}

 Accretion flows imbedding a shock wave have attracted much attention
 of researchers.
 Hydrodynamic instabilities of shocked accretion flows
 may explain the time variability of
 emissions from many black hole candidates,
 since a shock wave is a promising mechanism
 of transforming the gravitational energy into radiation.
 The possibility that a shock exists in black hole accretion disks
 was first suggested by \citet{Hawley84a,Hawley84b}
 even though they did not make clear the essential condition for its existence.
 The possible structure of the shocked accretion flows in equatorial plane 
 was described by \citet{Fukue87}, who suggested the importance of the
 multiplicity of critical points for the existence of the standing shock.
 Multiple critical points exist only for appropriate values
 of the injection parameters
 such as the specific angular momentum and Bernoulli constant.
 Note also that there are generally  two possible shock locations,
 which are referred to as the inner and outer shocks.

 The stability of the standing shock wave in the accretion flows
 has been also investigated by many authors
 both analytically and numerically.
 \citet{nak94,nak95} showed by linear analysis in the equatorial plane
 that if the post-shock matter
 are accelerated, the flow is unstable against radial perturbations,
 which is true for both Newtonian or general relativistic dynamics.
 As a result of this theorem,
 we know for the black hole accretion
 that the inner shock becomes generally unstable
 against radial perturbations
 and that non-rotating steady accretion flows to a black hole
 cannot have a stable standing shock wave in it.
 These features were also observed in 1D axisymmetric simulations
 for a pseudo-Newtonian potential \citep{cha93,nob94}.

 Recently, \citet{fog01,fog02} pointed out
 by linear analysis that the outer shock wave,
 which is stable against radial perturbations,
 is in fact unstable to non-radial axisymmetric perturbations.
 He argued that the advective-acoustic cycle could be
 responsible for the instability.
 In this mechanism the velocity and entropy fluctuations
 initially generated at the shock
 are advected inwards, producing pressure perturbations,
 which then propagate outwards, reach the shock and
 generate entropy and velocity fluctuations there,
 thus repeating the cycle with an increased amplitude.
 The same instability appears
 to be working in the accretion flows
 onto a nascent proto neutron star
 in the supernova core,
 in which large scale oscillations of the shock wave
 of $\ell=1$ nature are observed \citep{blo03},
 where $\ell$ stands for the index
 in the spherical harmonic functions $Y^{\ell}_{m}$.
 Although the mechanism is still controversial 
 \citep{ohn05,fog06sn,blomeza06,Lami07}, 
 the so-called standing accretion shock instability or SASI is
 currently attracting much attention
 as a promising explanation of asymmetric explosion of supernova
 as well as young pulsars' proper motions \citep{sch04,sch06}
 and spins \citep{blomeza07,Iwapre}.


 As for the stability of the shock wave
 against non-axisymmetric perturbations,
 \citet{molteni99} did 2D simulations of an adiabatic shocked accretion flow 
 by using the pseudo-Newtonian potential
 and found a non-axisymmetric instability.
 They showed that the shock instability is saturated at a low level
 and a new quasi-steady asymmetric configuration is realized.
 To investigate the mechanism of this instability,
 \citet{fog2003,Gu2005} performed
 linear analysis for both isothermal and adiabatic flows.
 They concluded that the instability
 seems to be of the Papaloizou-Pringle type
 and the repeated propagations of acoustic waves
 between the corotation radius and the shock surface are a driving mechanism.
 These conclusions are based on the WKB approximation
 and the comparison between the growth rate
 and the acoustic cycle period.

 It is also noted that
 \citet{yamasaki2007} investigated the linear stability
 of the shocked accretion flows onto the proto neutron star
 against non-axisymmetric perturbations in the equatorial plane.
 They demonstrated that the counter-rotating spiral modes are significantly damped,
 whereas the growth rate of the corotating modes
 is increased by rotation.
 They also claimed that the instability is
 not of the Papaloizou-Pringle type,
 since the stability is not affected
 by the presence or absence of the corotation point.
 Instead, they suggested the advective-acoustic cycle
 based on the WKB analysis.
 The purely acoustic cycles was found to be stable.

 As mentioned above, although many efforts have been made for clarifying
 the non-radial instability,
 the complete understanding of the mechanism is still elusive.
 It is noted that the unperturbed accretion flows should be treated properly,
 since they strongly affect the instability.
 In black hole accretions,
 the gravitation is one of the main factors to determine the flow features
 such as sonic points and shock locations.
 So far, however, the shock stability in the accretion flows to black holes
 has been investigated under the Newtonian
 or pseudo-Newtonian approximation and
 there has been no fully GR treatment.
 As shown later, the range of injection parameters that allows 
 the existence of a standing shock wave is changed substantially 
 when GR is fully taken into account.

 In this paper, we investigate fully general relativistically
 the stability of the shock in the advection-dominated accretion flows
 into a Schwarzschild black hole by using both
 linear analysis and non-linear dynamical simulations.
 In so doing, we consider only the equatorial plane,
 assuming that the $\theta$-component of four-velocity ($u^{\theta}$) and 
 all the $\theta$-derivatives are vanishing.
 We use the spherical coordinates ($r,\theta,\phi$) in the following.
 The evolution of the metric is not taken into account,
 which is justified if the mass of accretion flow is
 much smaller than the black hole mass.
 We show that the shock is indeed unstable against non-axisymmetric
 perturbations and a spiral arm structure is formed
 as the instability grows.
 We discuss the instability mechanisms
 comparing various time scales.
 Finally, we mention possible implications
 of our findings for gamma-ray bursts (GRBs)
 and black hole quasi-periodic oscillations (QPOs).

 This paper is organized as follows.
 In section 2, we describe
 the steady axisymmetric accretion flows with a shock in them.
 In section 3, we present the formulation of linear analysis.
 The numerical method for dynamical simulations
 is explained in section 4.
 The main numerical results and their analyses are given in section 5.
 The implication for GRBs
 and Black Hole QPOs are mentioned in section 6.
 Finally we summarize and conclude the paper in section 7.

\section{Axisymmetric Steady Accretion Flows with a Shock}
\subsection{The multiplicity of sonic points}
 One of the key features of the accretion flows to black holes
 is that the inflow velocity is supersonic at the event horizon.
 This immediately means that there should be at least two sonic points
 if a steady shock wave exists in the accretion flows,
 since both the pre- and post-shock flows are transonic.
 This is in sharp contrast to the accretions onto a neutron star,
 in which the post shock flow is subsonic.
 One of the consequences of this theorem is the fact that
 spherical adiabatic accretions into Schwarzschild black holes
 are unable to have a steady shock wave in them, since they have a
 single sonic point.
 As long as rotating accretion flows in the equatorial plane are concerned,
 the locations of these sonic points
 are determined by the adiabatic index and injection parameters,
 such as the Bernoulli constant and specific angular momentum.
 Note that the accretion rate is irrelevant for the locations
 of sonic points and standing shocks.

 The basic equations are the relativistic continuity equation and
 equation of energy-momentum conservations:
\begin{eqnarray}
\left(\rho_{0}u^{\mu}\right)_{;\mu} & = & 0 , \label{eq:continubase} \\
\left(T^{\mu\nu}\right)_{;\nu} & = & 0 , \label{eq:energymomentum}
\end{eqnarray}
 where the Greek indices represent the spacetime components.
 As already mentioned, we consider the accretions only in the equatorial plane,
 assuming $\theta$-component of velocity ($u^{\theta}$)
 and all $\theta$ derivatives are vanishing.
 Then the basic equations are reduced to
 ordinary differential equations with respect to the radial coordinate:
\begin{eqnarray}
&& \partial_{r} \left(r^2\rho_{0}u^r\right)
  =  0 , \label{eq:contibackground} \\
&& \partial_{r} p + \rho_0u^{r}\partial_{r}\left(hu_r\right)
 =  \frac{1}{2} \rho_0h \,
\biggl\{ \, \left( \partial_r g_{rr} \right) \left(u^r\right)^2 +
                \left( \partial_r g_{\phi\phi} \right) \left(u^{\phi}\right)^2 +
                \left( \partial_r g_{tt} \right) \left(u^t\right)^2 \, \biggr\} ,
 \label{eq:Eulerradialbackground} \\
&& \partial_{r}\left(hu_t\right) =  0 , \label{eq:enmotimebackground} \\
&& \partial_{r}\left(hu_{\phi}\right) =  0 . \label{eq:enmophibackground}
\end{eqnarray}
 Note that these treatments are slightly different from those by
 \citet{molteni99,fog2003,Gu2005}.
 They employed the cylindrical coordinates
 and integrated out the vertical structures, thus considering the 
 accretion flows only in the equatorial plane.
 On the other hand, we use the spherical coordinates in this paper, 
 since it is mathematically more convenient
 for the fully general relativistic approach.
 Because of this difference, however,
 the obtained accretion flows would be different from the ones
 considered in this paper
 if the general relativity were taken into account
 in their formulations.

\begin{figure}
\includegraphics[angle=0,scale=.45]{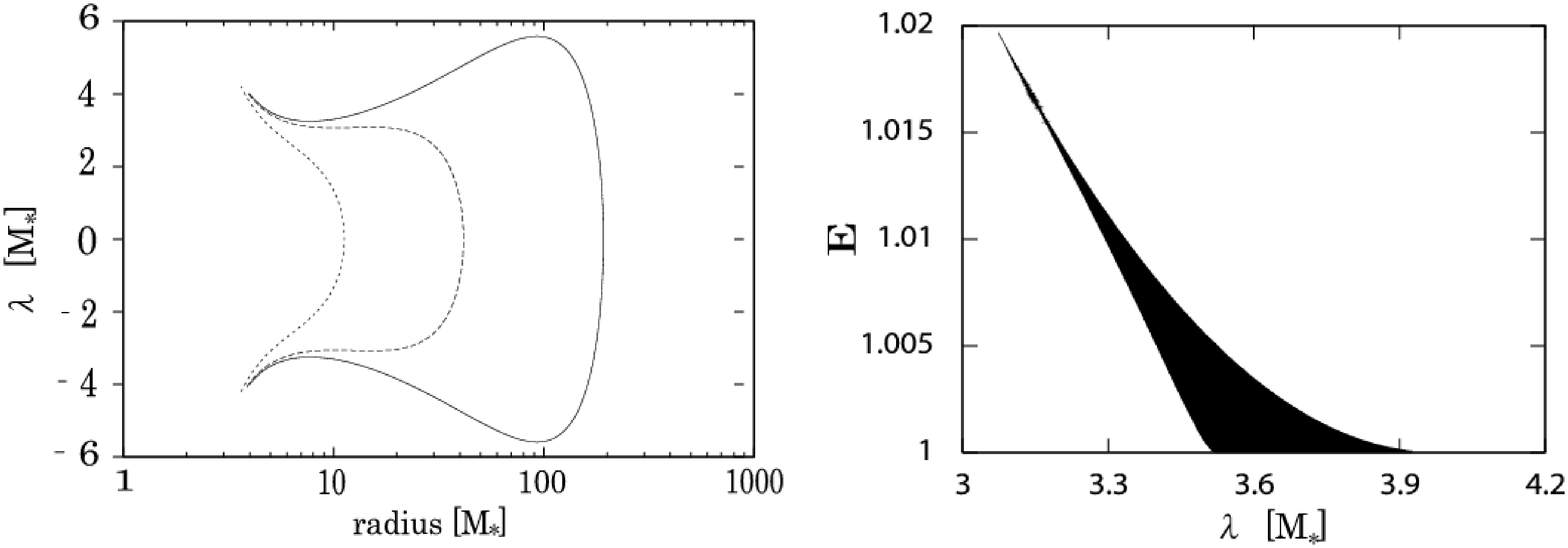}

\caption{Left: The location of sonic points
 as a function of the Bernoulli constant ($E$)
 and specific angular momentum ($\lambda$)
 in the Schwarzschild geometry.
 The solid, dashed, dotted lines correspond to
 $E = 1.004$, $1.02$ and $1.1$, respectively.
 Right: The injection parameters 
 for the existence of a standing shock wave. The shaded region
 allows the standing shock wave.
 The adiabatic index is $4/3$ for both panels.}
\label{tefig1}
\end{figure}

 The left panel of Figure~\ref{tefig1} shows
 the locations of the sonic points in Schwarzschild black hole
 as a function of the Bernoulli constant
 and specific angular momentum (see also \citet{lu986}).
 As shown in the figure, there are indeed two or three sonic points
 for some combinations of
 the Bernoulli constant and angular momentum.
 The maximum and minimum specific 
 angular momenta are $\lambda_{max}\sim 4.0M_{\ast}$ and $\lambda_{min}\sim 3.2M_{\ast}$,
 respectively, for the Bernoulli constant $E=1.004$, for example.
 It should be noted that one of the most important differences
 between the full GR and the pseudo-Newtonian treatments
 is the range of the injection parameters that allows the existence of
 multiple sonic points. As the Bernoulli constant becomes smaller,
 the maximum specific angular momentum gets larger without limit for the
 pseudo Newtonian case whereas it is bounded for the full GR case.
 For the Bernoulli constant that may be typical for massive stellar collapse,
 e.g. $E=1.003$, the maximum specific angular momentum is larger  by about 60\% 
 for the pseudo Newtonian approximation than for the full GR treatment.
 The reason for this difference is that the gravity nearby the black hole 
 is too strong in the pseudo-Newtonian approximation. It should be pointed out 
 that the region of the injection parameters that allows multiple sonic points is 
 not very wide and the above-mentioned difference may be important in considering 
 the implications for astrophysical phenomena.

\subsection{The locations of standing shock waves}

 The sonic points discussed in the previous subsection
 correspond to the so-called critical points in dynamical system.
 It is known that the innermost and outermost critical points are
 of the saddle-type, while the middle critical point is of the
 center-type.
 Hence, the transonic accretion flows can be constructed only for the
 former two critical points.
 These two transonic flows have the same Bernoulli constant
 and angular momentum but different entropies,
 and they can be connected by a standing shock wave, where
 the Rankine-Hugoniot relations hold:
\begin{eqnarray}
 [ \, \rho_0 u^{\mu}  \,] \, l_{\mu} & = & 0 ,
 \label{eq:RHconti}
\end{eqnarray}
\begin{eqnarray}
 [ \, T^{\mu \nu} \,] \, l_{\nu} & = & 0 ,
 \label{eq:RHem}
\end{eqnarray}
 where $l_{\mu}$ is a 4-dimensional vector normal to the shocked surface,
 and is set to be $l_{\mu}=\left(0,1,0,0\right)$ in axisymmetric steady flows
 (see also Eq. (\ref{eq:normalvectoshock})
 for the non-axisymmetric perturbations).
 We use the notation of
 $[\, Q \, ] \equiv Q_{+} - Q_{-}$,
 where the subscript $(+)$ represents a post-shock quantity
 and $(-)$ a pre-shock quantity.
 The Bernoulli constant and specific angular momentum
 are defined respectively as
\begin{eqnarray}
 E &\equiv& -hu_{t}, \\
 \lambda &\equiv& - \frac{u_{\phi}}{u_{t}}.
\end{eqnarray}
Since the Bernoulli constant, specific angular momentum and mass flux
 are unchanged across the shock,
 we only need to consider the radial component of
 energy momentum conservation across the shock.
 Then the Rankine-Hugoniot relation in axisymmetric steady flows
 can be written as:
\begin{eqnarray}
[\, \rho_0 h u_{r} u^r + p \,] & = & 0.
 \label{eq:RHenmostaaxis}
\end{eqnarray}

 In general, there are two possible shock locations,
 to which we refer as the inner and outer shocks.
 This is apparent in the right panel of the Figures~\ref{tefig1},
 in which we show the parameter region that allows 
 multiple shock locations.
 It is well known, however, that the inner shock is
 unstable against axisymmetric perturbations
 \citep{nak95}, which we have confirmed by our own linear analysis.
 We have also done numerical simulations
 with a single grid point in the azimuthal direction,
 thus suppressing non-axisymmetric modes, and observed
 that the inner shock is either swallowed by a black hole or moved outwards
 to be the outer shock after radial perturbations are imposed.
 On the other hand, we have seen that the outer shock is stable
 against radial perturbations even if the perturbation amplitude
 is not necessarily small.
 In the following, we will consider only the outer shock.

 We have constructed several axisymmetric steady accretion flows
 with an outer shock for different combinations
 of the adiabatic index and injection parameters,
 which are summarized in Table~\ref{tab1}.

\section{Linear Analysis of non-axisymmetric shock instability}
 Here we give the basic equations and boundary conditions
 for the linear analysis.
 The obtained eigen values are
 later compared with the numerical simulations,
 and the eigenstates are employed
 to impose the initial perturbations.

 The basic equations are the linearized relativistic continuity and
 energy-momentum tensor conservation equations
 (see Eqs. (\ref{eq:continubase}) and (\ref{eq:energymomentum})).
 We again neglect the $\theta$-component of velocity
 and all the $\theta$-derivatives and consider the equatorial plane only.
 Under this assumption, the system equations are written as follows:
\begin{eqnarray}
\partial_rf & = & 
\frac{i}{\rho_{0(0)} u^{r(0)}}\biggl\{ \,\rho_{0(1)}u^{t(0)}\sigma +
                       \rho_{0(0)}\left( \omega u^{t(1)}-mu^{\phi(1)} \right)\, \biggr\} ,
 \label{eq:lineconti2} \\
\partial_rq & = & \frac{i}{\rho_{0(0)}u^{r(0)}h_{(0)}u^{t(0)}}
 \left(\, \omega p_{1} + \rho_{0(0)} u^{t(0)} h_{(0)} u_{t(0)}
 \sigma q \,\right) , \label{eq:ber2}\\
\partial_rV_{(1)} & = & 
i \frac{u^{t(0)}}{u^{r(0)}} \sigma V_{(1)} , \label{eq:linework2} \\
\partial_rS_{(1)} & = & 
i \frac{u^{t(0)}}{u^{r(0)}} \sigma S_{(1)} , \label{eq:lineentro2}
\end{eqnarray}
where $S$ is entropy and the following notations are used:
\begin{eqnarray}
f & \equiv & \frac{\rho_{0(1)}}{\rho_{0(0)}} + \frac{u^{r(1)}}{u^{r(0)}} , \label{eq:fdef} \\
q & \equiv & \frac{h_{(1)}}{h_{(0)}} + \frac{u_{t(1)}}{u_{t(0)}} , \label{eq:qdef} \\
V_{(1)} & \equiv & \omega\left(\, h_{(1)} u_{\phi(0)} + h_{(0)}u_{\phi(1)} \,\right)
+ m\left(\, h_{(1)} u_{t(0)} + h_{(0)} u_{t(1)} \,\right) , \label{eq:Vdef} \\
\sigma & \equiv & \omega - m \frac{u^{\phi(0)}}{u^{t(0)}} . \label{eq:sigmadef}
\end{eqnarray}
 Following the standard procedure of linear stability analysis,
 the perturbed quantities are assumed to be proportional to $e^{-i\omega t + im\phi}$.
 All perturbed quantities are calculated from $ f , q , V_{(1)} $ and $ S_{(1)} $.
 $V_{(1)}$ and $ S_{(1)}$ can be integrated analytically as
\begin{eqnarray}
V_{(1)} & = & V_{(1)} \biggl|_{R} \,
 \exp \left( \, i \frac{u^{t(0)}}{u^{r(0)}} \sigma \, \right) ,
\label{eq:workinteg} \\
S_{(1)} & = & S_{(1)} \biggl|_{R} \,
 \exp \left( \, i \frac{u^{t(0)}}{u^{r(0)}} \sigma \, \right) ,
\label{eq:entrointeg} 
\end{eqnarray}
 where the subscript 'R' denotes the value evaluated at $r=R$.
 Thus, we need to integrate numerically
 only two Eqs. (\ref{eq:lineconti2}) and (\ref{eq:ber2}).

 These linearized equations can be solved
 with appropriately setting boundary conditions,
 which are imposed at the shock surface and inner sonic point.
 In this study, we assume that the perturbations
 are confined in the post-shock region and
 the pre-shock region remains unperturbed.
 As a result, the outer boundary condition is set at the shock surface.
 We express the shock radius as follows:
\begin{eqnarray}
 R_{sh} = R_{sh(0)} +
 \eta \exp\left(\,-i \omega t + i m \phi \,\right) ,
 \label{eq:Rshpertu}
\end{eqnarray}
 where $\eta$ denotes the initial amplitude of shock displacement.
 Defined as the 4-dimensional vector normal to the shock surface,
 $l_{\nu}$ can be given as
\begin{eqnarray}
 l_{\nu} =
 \left( i \omega \eta e^{-i \omega t + i m \phi }, 1 , 0 , - i m \eta e^{-i \omega t + i m \phi } \right) .
 \label{eq:normalvectoshock}
\end{eqnarray}
 Using these relations in the Rankine-Hugoniot relations expressed in general as
 $[ \, Q^{\mu}  \, ]  \, l_{\mu} = 0  $,
 we can write Eqs. (\ref{eq:RHconti}) and (\ref{eq:RHem}) as
\begin{eqnarray}
&&  [ \, Q^{t(0)} \,] i \omega \eta 
- [ \, Q^{\phi(0)} \,] i m \eta
+ [ \,Q^{r(1)} \,]  =  0 , \label{eq:geneRH}
\end{eqnarray}
where the following notations are employed:
\begin{eqnarray}
&& Q^{\mu(0)}  \equiv  Q^{\mu(0)} \biggl|_{R_{sh(0)}} , \label{eq:geneRHunp} \\
&& Q^{\mu(1)}  \equiv  Q^{\mu(1)} \biggl|_{R_{sh(0)}}
+ \eta \left( \frac{d}{dr} Q^{\mu(0)} \right) \biggl|_{R_{sh(0)}} . \label{eq:geneRHunppertu}
\end{eqnarray}
 As mentioned above,
 we assume that the pre-shock quantities are unperturbed,
 i.e $ \left(Q^{\mu(1)}|_{R_{sh(0)}} \right)_{-}=0$.
 The explicit forms of these equations are given by
\begin{eqnarray}
 \left( \rho_{0(0)} u^{r(1)} \right)_{+} + \left( \rho_{0(1)} u^{r(0)} \right)_{+} + A = 0 ,
 \label{eq:lineshocksurface1} \\
 \biggl\{ \left(\rho_0 h u^t u^r\right)^{(0)} \, 
    \left( \frac{\rho_{0(1)}}{\rho_{0(0)}} 
    + \frac{h_{(1)}}{h_{(0)}}
    + \frac{u^{t(1)}}{u^{t(0)}}
    + \frac{u^{r(1)}}{u^{r(0)}}       \right) \biggr\}_{+} + B = 0 ,
 \label{eq:lineshocksurface2} \\
\biggl\{  \left( \rho_0 h \left(u^r\right)^2 \right)^{(0)} \,
    \left( \frac{\rho_{0(1)}}{\rho_{0(0)}} 
    + \frac{h_{(1)}}{h_{(0)}}
    + 2 \frac{u^{r(1)}}{u^{r(0)}} \right) + p_{(1)} g^{rr}\biggr\}_{+} + C = 0 ,
 \label{eq:lineshocksurface3} \\
 \biggl\{ \left(\rho_0 h u^{\phi} u^r\right)^{(0)} \, 
    \left( \frac{\rho_{0(1)}}{\rho_{0(0)}} 
    + \frac{h_{(1)}}{h_{(0)}}
    + \frac{u^{\phi(1)}}{u^{\phi(0)}}
    + \frac{u^{r(1)}}{u^{r(0)}}       \right) \biggr\}_{+} + D = 0 ,
 \label{eq:lineshocksurface4}
\end{eqnarray}
with the following definitions:
\begin{eqnarray}
 A &\equiv& \biggl\{ \left( \frac{d}{dr} \rho_{0(0)} u^{r(0)} \right)_{+}
            - \left( \frac{d}{dr} \rho_{0(0)} u^{r(0)} \right)_{-} \biggr\}
            \, \eta
          + i \omega \eta \,
          \biggl\{ \left( \rho_{0(0)} u^{t(0)} \right)_{+} -
             \left( \rho_{0(0)} u^{t(0)} \right)_{-}  \biggr\} \nonumber \\
&&          - i m \eta \,
          \biggl\{ \left( \rho_{0(0)} u^{\phi(0)} \right)_{+} -
             \left( \rho_{0(0)} u^{\phi(0)} \right)_{-}  \biggr\} ,
          \label{defworkA} \\
 B &\equiv& \biggl\{ \left( \frac{d}{dr} \rho_{0(0)} h_{0} u^{t(0)} u^{r(0)} \right)_{+}
              - \left( \frac{d}{dr} \rho_{0(0)} h_{0} u^{t(0)} u^{r(0)} \right)_{-} \biggr\}
              \, \eta
 \nonumber \\
&&          + i \omega \eta \,
          \biggl\{  \left(\rho_{0} h \left( u^t \right)^2 + p g^{tt}\right)^{(0)}_{\!+} -
              \left(\rho_{0} h \left( u^t \right)^2 + p g^{tt}\right)^{(0)}_{\!-}  \biggr\} 
 \nonumber \\
&&          - i m \eta \,
          \biggl\{  \left(\rho_{0} h u^t u^{\phi}\right)^{(0)}_{\!+} -
              \left(\rho_{0} h u^t u^{\phi}\right)^{(0)}_{\!-}  \biggr\} ,
          \label{defworkB} \\
 C &\equiv& \biggl\{ \left( \frac{d}{dr}
               \left( \rho_{0(0)} h_{(0)} \left(u^{r(0)}\right)^{2}
                      + p^{(0)} g^{rr} \right)\right)_{+}
              -\left( \frac{d}{dr}
               \left( \rho_{0(0)} h_{(0)} \left(u^{r(0)}\right)^{2} + p^{(0)} g^{rr} \right)\right)_{-} \biggr\}
             \, \eta
 \nonumber \\
&&          + i \omega \eta \,
          \biggl\{  \left(\rho_{0} h u^t u^r \right)^{(0)}_{\!+} -
              \left(\rho_{0} h u^t u^r \right)^{(0)}_{\!-}  \biggr\} 
 \nonumber \\
&&          - i m \eta \,
          \biggl\{  \left(\rho_{0} h u^r u^{\phi}\right)^{(0)}_{\!+} -
              \left(\rho_{0} h u^r u^{\phi}\right)^{(0)}_{\!-}  \biggr\} ,
          \label{defworkC} \\
 D &\equiv& \biggl\{ \left( \frac{d}{dr} \rho_{0(0)} h_{0} u^{\phi(0)} u^{r(0)} \right)_{+}
             - \left( \frac{d}{dr} \rho_{0(0)} h_{0} u^{\phi(0)} u^{r(0)} \right)_{-} \biggr\}
              \, \eta
 \nonumber \\
&&          + i \omega \eta \,
          \biggl\{  \left(\rho_{0} h u^{\phi} u^t \right)^{(0)}_{\!+} -
              \left(\rho_{0} h u^{\phi} u^t \right)^{(0)}_{\!-}  \biggr\} 
 \nonumber \\
&&          - i m \eta \,
          \biggl\{  \left(\rho_{0} h \left(u^{\phi}\right)^2 + p g^{\phi \phi}\right)^{(0)}_{\!+} -
              \left(\rho_{0} h \left(u^{\phi}\right)^2 + p g^{\phi \phi}\right)^{(0)}_{\!-}  \biggr\} .
          \label{defworkD}
\end{eqnarray}

 As for the inner boundary condition,
 on the other hand, a regularity condition is imposed at the sonic point.
 By combining Eqs. (\ref{eq:lineconti2}) to (\ref{eq:sigmadef}),
 we obtain the differential equation for $u_{r}$,
 which is written generally as
\begin{eqnarray}
  F \left(u_r\right)_{,r}  = G . \label{eq:difaboutudnr}
\end{eqnarray}
 The linearized form of this equation becomes
\begin{eqnarray}
 \left(u_{r(1)}\right)_{,r} = \frac{G_{(1)} - F_{(1)}u_{r(0),r}}{F_{(0)}} ,
\end{eqnarray}
where the explicit forms of $F_{(0)}$, $F_{(1)}$ and $G_{(1)}$ are
\begin{eqnarray}
F_{(0)} & \equiv & \rho_{0(0)} h_{(0)} g^{rr}
\biggl\{  u^{r(0)} u_{r(0)} - \left(b_{s(0)} \right)^2 \left( 1 + u^{r(0)} u_{r(0)} \right)  \biggr\}
 \label{eq:Fbackground} \\
F_{(1)} & \equiv & \, 
   \rho_{0(1)} h_{(0)} \left(u^{r(0)}\right)^2
 + \rho_{0(0)} h_{(1)} \left(u^{r(0)}\right)^2
 + 2 \rho_{0(0)} h_{(0)} u^{r(0)} u^{r(1)} \nonumber \\
&&  \, 
 - \Gamma g^{rr} \biggl\{ p_{(1)} \left( 1+u^{r(0)} u_{r(0)} \right)
                   + 2 p_{(0)} u^{r(1)} u_{r(0)} \biggr\} ,
   \label{eq:woregso2} \\
 G_{(1)} & \equiv & \, \Gamma \left( g^{rr} \! _{,r} + \frac{2}{r} \right)
 \biggl\{ \left( p_{(1)} u_{r(0)} + p_{(0)} u_{r(1)} \right)\left( 1 + u_{r(0)} u^{r(0)} \right)
   + 2 p_{(0)} \left( u_{r(0)} \right)^2 u^{r(1)}  \biggr\} \nonumber \\
 &&  + \, \frac{1}{2} \biggl\{ \rho_{0(1)} h_{(0)} u^{r(0)} 
                   + \rho_{0(0)} h_{(1)} u^{r(0)}
                   + \rho_{0(0)} h_{(0)} u^{r(1)} \biggr\} 
    \biggl\{  g_{rr,r} \left(u^{r(0)}\right)^2 
      + g_{\phi \phi ,r} \left(u^{\phi(0)}\right)^2
      + g_{ t t ,r } \left(u^{t(0)}\right)^2  \biggr\}  \nonumber \\
 &&  + \,  \rho_{0(0)} h_{(0)} u^{r(0)}
      \biggl\{ g_{rr,r} u^{r(0)} u^{r(1)} +
         g_{\phi \phi ,r} u^{\phi(0)} u^{\phi(1)} +
         g_{tt,r} u^{t(0)} u^{t(1)} \biggr\}   \nonumber \\
 &&  + \, i \rho_{0(0)} h_{(0)} u^{r(0)} u^{t(0)} u_{r(1)} \sigma \nonumber \\
 &&  + \,  i \Gamma p_{(0)} \left( 1 + u_{r(0)} u^{r(0)} \right)
                       \left( - \omega u^{t(1)} + m u^{\phi (1)} \right) \nonumber \\
 &&  -\,  i u^{t(0)} p_{(1)} \sigma  . \label{eq:woregso1}
\end{eqnarray}
 $b_{s(0)}$ and $\Gamma$ are the unperturbed sound velocity
 in the comoving frame and
 the adiabatic index, respectively.
 Since $F_{(0)}$ vanishes at the sonic point,
 we impose the regularity condition there:
\begin{eqnarray}
 G_{(1)} - F_{(1)}u_{r(0),r} = 0 . \label{eq:regso}
\end{eqnarray}
\section{Numerical Method and Models}

 The growth of the initial perturbations
 is computed with a multi-dimensional general 
 relativistic hydrodynamics code, which is based on a
 modern technique, the so-called high-resolution central scheme
 \citep{kur00}.
 The details of the numerical scheme are
 given in Appendix.

 We use the Kerr-Schild coordinates with
 the Kerr parameter being set to be zero,
 since they have no coordinate singularity at the event horizon
 and we can put the inner boundary inside the event horizon.
 This is very advantageous for numerical simulations.
 We employ a $\Gamma$-law~EOS,
 $p = \left(\Gamma-1\right)\rho_0 \epsilon$,
 where $p$ and $\rho_0$ and $\epsilon$ are the pressure,
 rest-mass density, and specific internal energy, respectively.

 The computational domain is a part of the equatorial plane with
 $ 1.5M_{\ast}  \leq r \leq 200M_{\ast}$ and the computation times are 
 $ \sim 6 \times 10^4 M_{\ast}$,
 where $M_{\ast}$ denotes the black hole mass
 and the unit with $c=G=1$ is used.
 We employ $600(r) \times 60(\phi)$ grid points.
 The radial grid width is non-uniform
 with the grid being smallest, ($ \Delta r = 0.1 M_{\ast}$),
 at the inner boundary and increasing
 geometrically by $0.34 \%$ per zone toward the outer boundary.

 The initial perturbation modes
 are summarized in Table~\ref{tab1}.
 For Models M1 to M12
 we add the $m=1$ mode,
 where $m$ stands for the azimuthal mode number in $e^{im\phi}$.
 For Models M1m2 and M1m3, on the other hand,
 the $m=2,3$ modes are initially imposed, respectively.
 These models are meant for the study of the
 initial-mode dependence.
 In order to investigate the initial-amplitude dependence,
 we also run Models M1a10 and M1a100,
 whose initial amplitudes are $10 \%$ and $100 \%$, respectively.
 The unperturbed flows for Models M1m2, M1m3, M1a10 and M1a100
 are common to those of the other models.

 The radial distributions of these modes
 are obtained by the linear analysis.
 This is important to compare
 the linear growth of purely single modes
 between the linear analysis and numerical simulations,
 and is one of the differences from \citet{molteni99}.
 We choose the most unstable mode
 for each azimuthal wave number except for M1a100,
 in which the following perturbation is employed
 to avoid a velocity larger than the light velocity:
\begin{equation}
 v^r = v^r \! _{sta}
 \biggl\{  1 + \sin \left( \phi \right)  \biggr\},
\end{equation}
 where $v^r \! _{sta}$ is the unperturbed radial velocity.
 The initial perturbations are added
 to the whole post-shock region,
 except for Model M1a100,
 in which the perturbation is imposed to the region
 between the inner sonic point and the shock again
 to prevent the flow velocity
 from being larger than the light velocity.


 In the following, we set the black hole mass
 to be $M_{\ast}=3$M$_{\sun}$,
 where $M_{\sun}$ is the solar mass.
 We have in mind here some applications
 to astrophysical phenomena such as GRBs and QPOs.
 Since the formulation is dimensionless,
 the scaling is quite obvious.

\section{Numerical Results by GRHD}
 In this section, we describe
 the time evolutions of non-axisymmetric instability
 obtained by fully dynamical GRHD simulations.
 In the following analysis,
 we frequently employ the mode decomposition
 of the shock surface by the following Fourier transform:
\begin{equation}
 a_m\left(t\right) = \int_0^{2\pi} R_{sh}\left(\phi,t\right) \,
 e^{im\phi} \, d\phi ,
\end{equation}
 where $R_{sh}\left(\phi,t\right)$ and $a_m\left(t\right)$ are
 the radius of the shock wave as a function of $\phi$ and $t$
 and the amplitude of mode $m$ as a function of $t$, respectively.
\subsection{Basic Features}
\begin{figure}
\includegraphics[angle=0,scale=.20]{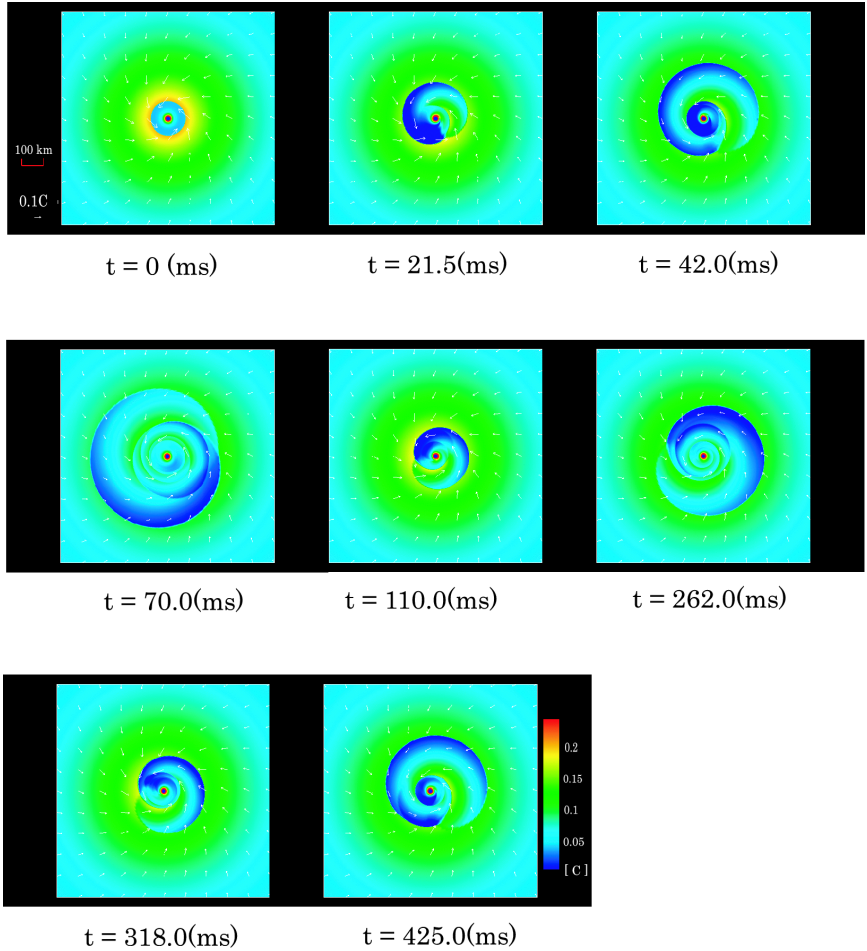}

\caption{The time evolution of velocity for Model M1.
 The color contour shows the magnitude of radial velocity.
 The arrows represent the velocities at their positions.
 The central region in blue corresponds to the black hole.}
\label{tefig5}
\end{figure}

\begin{figure}
\includegraphics[angle=0,scale=.20]{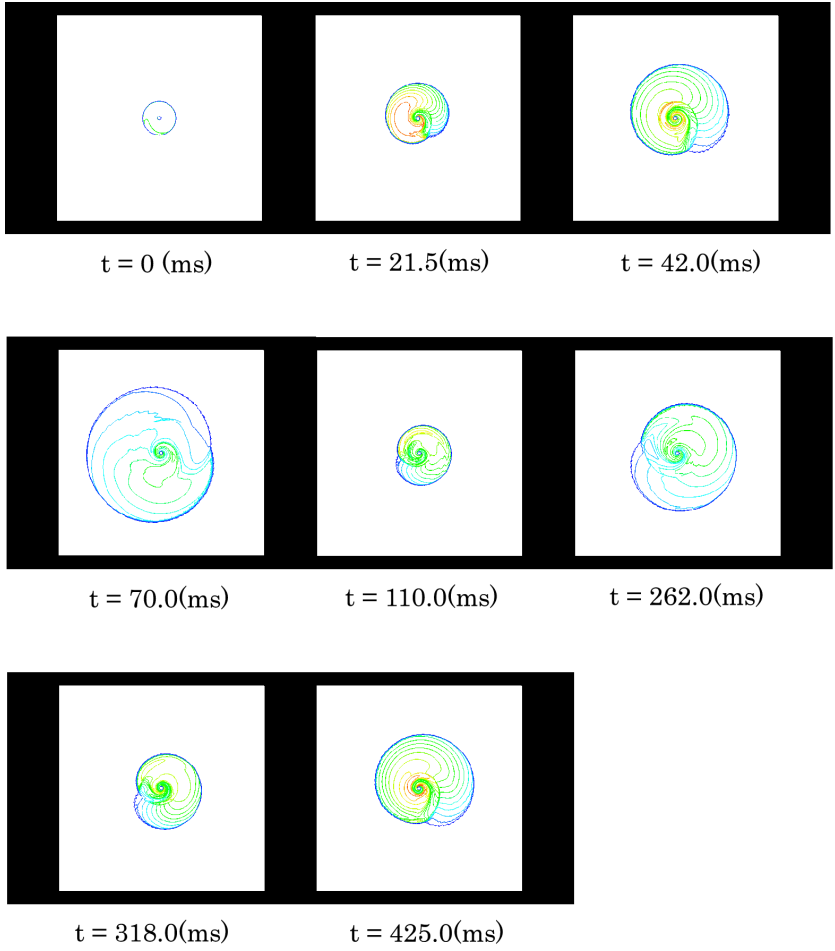}

\caption{The time evolution of entropy for Model M1.}
\label{tefig6}
\end{figure}

 Figures~\ref{tefig5} and \ref{tefig6} show
 the temporal evolutions of velocity and entropy
 for the baseline model M1, respectively.
 The perturbation grows exponentially at the beginning and
 the shock wave is deformed according to the imposed $m=1$ mode,
 for which the deformed shock surface rotates progressively,
 that is, in the direction of the unperturbed flows.
 Then the shock radius, or the $m=0$ mode, starts to grow.
 After several revolutions,
 a spiral arm develops and
 the instability is saturated with
 more complex structures.
 In this non-linear regime,
 several shocks are generated and collide with each other.
 As a result of these interactions,
 the original shock oscillates radially.

\begin{figure}
\includegraphics[angle=0,scale=.50]{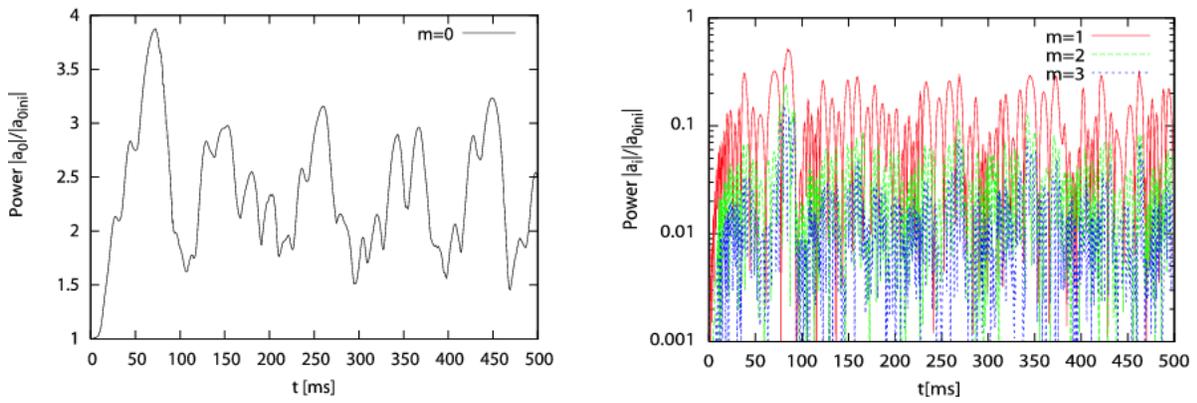}

\caption{The time evolutions of the $m=0$ mode (left) and
 the $m=1,2,3$ modes (right).}
\label{figM1modet500}
\end{figure}

 As given in Table~\ref{tab:dyreso1},
 the saturation levels of shock radius, or the $m=0$ mode,
 differ widely among models.
 For example, the left panel of Figure~\ref{figM1modet500}
 shows the time evolution of $m=0$ mode for Model M1.
 Both large- and small-amplitude oscillations can be seen,
 which correspond to the periods of $\sim100$~ms and $\sim20$~ms, respectively.
 These two kinds of axisymmetric oscillations
 are also found in other models.
 Their characteristics are summarized
 in Table~\ref{tab:dyreso1}.
 The growth of $m=0$ mode in the non-linear phase
 is strongly dependent on the mach number
 (see Tables~\ref{tab1}~and~\ref{tab:dyreso1} ).
 i.e., a stronger shock tends to become more unstable
 against non-axisymmetric perturbations. Although there 
 is no entry to the maximum amplitude of m=0 mode in Table 2 
 for Model M5, this is not an exception. In fact, 
 the shock is so unstable in this model that it leaves 
 the computational domain soon after the addition of 
 non-axisymmetric perturbations.
 
 On the other hand, the oscillation periods of the non-axisymmetric modes
 are much shorter in general
 than those of the axisymmetric ones.
 Indeed, the right panel of Figure~\ref{figM1modet500}
 shows the time evolutions of $m=1,2,3$ modes for Model M1.
 The typical periods range from several to a few dozen milli-seconds.

 In the non-linear phase,
 the dominant mode is $m=1$ for almost all models.
 Remarkably, although the initially added perturbations
 are not the $m=1$ mode in Models M1m2 and M1m3,
 the non-linear mode couplings lead eventually
 to the dominance of the $m=1$ mode
 (see the upper panels of Figure~\ref{figmodedepegraph}).

\subsection{Comparison with linear analysis}
\begin{figure}
\includegraphics[angle=0,scale=.60]{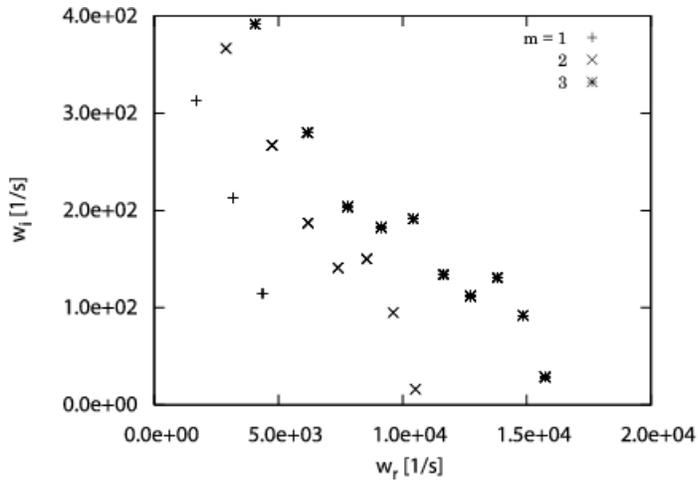}

\caption{The real and imaginary parts of eigen frequencies
 for some of the $m=1,2,3$ modes for Model M1.}
\label{figlinem123eigen}
\end{figure}

 Equations (\ref{eq:lineconti2}) to (\ref{eq:lineentro2})
 with the boundary conditions at the shock surface and sonic point
 are solved numerically to find eigen modes.
 Figure~\ref{figlinem123eigen} shows the real and imaginary parts
 of eigen frequencies for some of the $m = 1 , 2 , 3$ modes
 for Model M1.
 They are all unstable non-axisymmetric modes.
 We find, on the other hand, that the axisymmetric perturbations
 are stable, which is consistent with both the present dynamical
 simulations and the previous work \citep{nak95}.
 The oscillation periods, which correspond to $\omega_r$,
 are $1.5\sim3.7$~ms for the most unstable mode
 in each m-sequence for Model M1,
 whereas the growth times,
 which are obtained from $\omega_i$, are
 $2.6\sim3.2$~ms.

\begin{figure}
\includegraphics[angle=0,scale=.60]{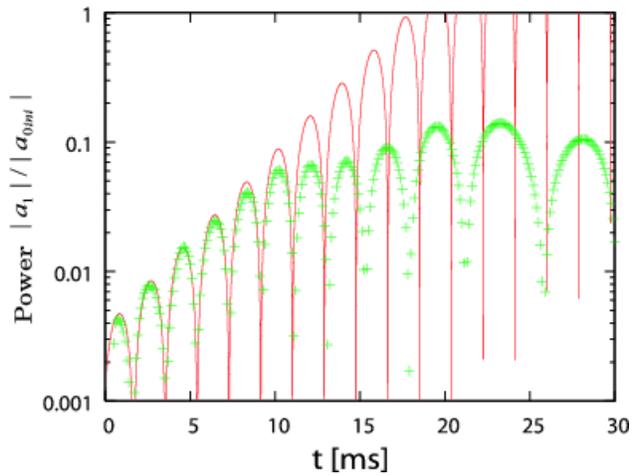}

\caption{The comparison of the time evolutions of the amplitudes of
 $m=1$ mode for Model M1 obtained by the
 linear analysis and dynamical simulation.
 The red line shows the evolution expected by the linear analysis,
 while the green crosses are the simulation results.}
\label{figcomplineGRHD}
\end{figure}

\begin{figure}
\includegraphics[angle=0,scale=.60]{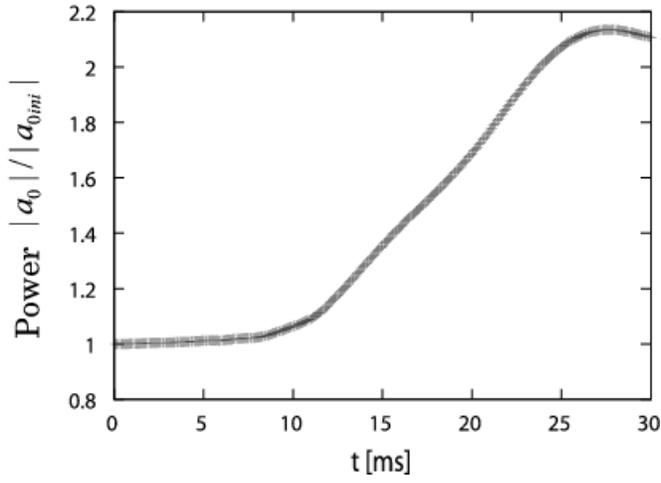}

\caption{The evolution of the amplitude of $m=0$ mode
 for Model M1.}
\label{figM1m0evo}
\end{figure}

 Figure~\ref{figcomplineGRHD} shows the comparison
 of the amplitudes obtained by the linear analysis
 and dynamical simulations.
 As can be clearly seen,
 the oscillation period and growth time
 are in good agreement between them for the first $10$~ms,
 which is the linear growth phase.
 It is also found
 that after $10$~ms, the result of the dynamical simulation
 starts to deviate from that of the linear analysis,
 which indicates the beginning of the non-linear phase.
 The amplitude is saturated and the oscillation period
 gets slightly longer.
 According to Figure~\ref{figM1m0evo},
 which shows the evolution of axisymmetric $m=0$ mode,
 we find that it starts to grow at $t\sim10$~ms,
 leading to the increase of the average shock radius.
 This is the reason why the oscillation period
 becomes longer in the non-linear phase.


 As mentioned in the previous section
 and summarized in Table~\ref{tab:dyreso1},
 the dominant modes in the non-linear phase are
 progressive, that is,
 the deformation pattern rotates in the same direction
 as the unperturbed flow.
 This is also true for all the linearly unstable modes.
 In fact, the linear analysis shows that
 there is no unstable mode for $ m < 0$.
 Note that \citet{yamasaki2007} demonstrated by linear analysis
 for the accretion onto a neutron star that the progressive modes
 are enhanced and retrogressive ones are suppressed by rotation.

 As the specific angular momentum of the unperturbed flow becomes larger,
 the distance between the shock wave and the inner sonic point
 gets greater (See Models M1 and M4 in Table~\ref{tab1}).
 According to the linear analysis,
 the number of unstable modes also increases,
 whereas the growth time of the most unstable modes becomes longer.
 This suggests that the larger distance tends to
 stabilize the non-axisymmetric instability
 though it is not the only factor
 for the shock instability.
 In fact, it is also found that the growth rate of unstable modes
 is affected by the shock strength, that is,
 stronger shocks tend to be more unstable.

\begin{figure}
\includegraphics[angle=0,scale=.50]{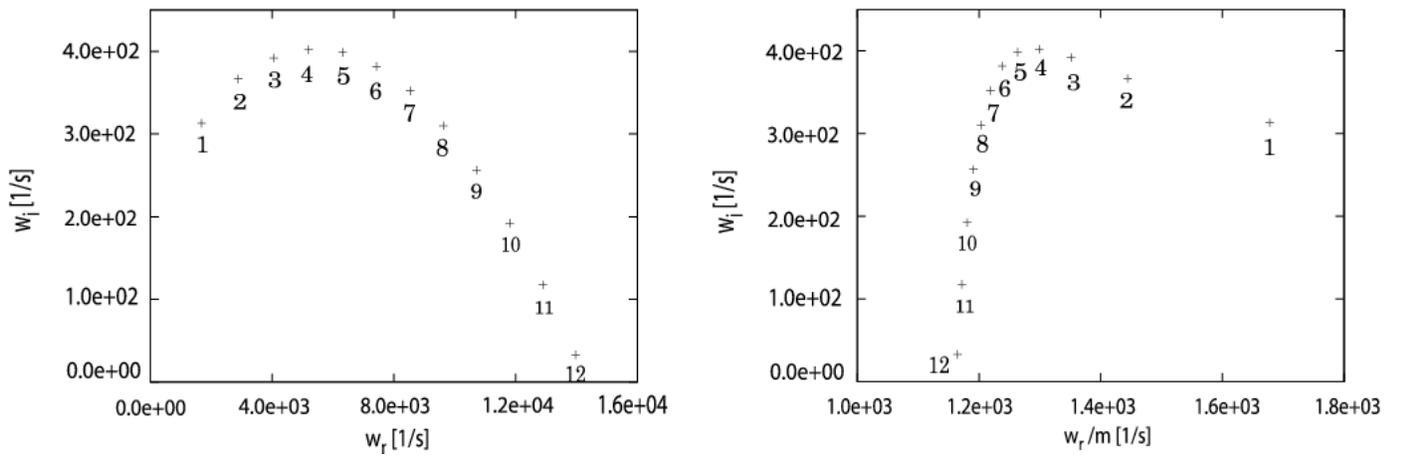}

\caption{Left: the most unstable eigenfrequency in each mode.
 Right: the same as in left figure
 but the horizontal axis is the pattern frequency $\omega_r / m$}
\label{fig:mostunstamodesM1case}
\end{figure}

 Finally we show in Figure~\ref{fig:mostunstamodesM1case}
 the most unstable modes for different $m$'s in Model M1.
 As is clear, the most unstable of all is the $m=4$ mode
 and the modes with $m > 12$ are stable for this model.
 It is also interesting to note that
 the real part of eigen frequency
 for the most unstable modes
 becomes larger as the mode number $m$ is higher
 (see the left panel of Figure~\ref{fig:mostunstamodesM1case}),
 while the pattern frequency $\omega_r/m$ becomes smaller
 (the right panel of Figure~\ref{fig:mostunstamodesM1case}).

\subsection{Dependence on the initial perturbations}
 By comparing Models M1, M1a10 and M1a100,
 we find that the qualitative feature
 of dynamics are almost the same.
 In the linear phase,
 the growth of mode amplitudes is unaffected by the
 initial condition.
 Only the duration of the linear phase become shorter
 as the initial amplitude is larger as expected.
 The saturation levels do not differ very much among three models.
 In fact, the non-linear phase is rather chaotic
 and forgets the difference in the initial condition.
 It is also important to point out that
 in spite of the large initial amplitude for Model M1a100,
 the shock continues to exist.
 This implies that if the injection parameters are appropriate,
 a standing shock will exist oscillating violently
 in the accretion flows into black holes.

\begin{figure}
\includegraphics[angle=0,scale=.40]{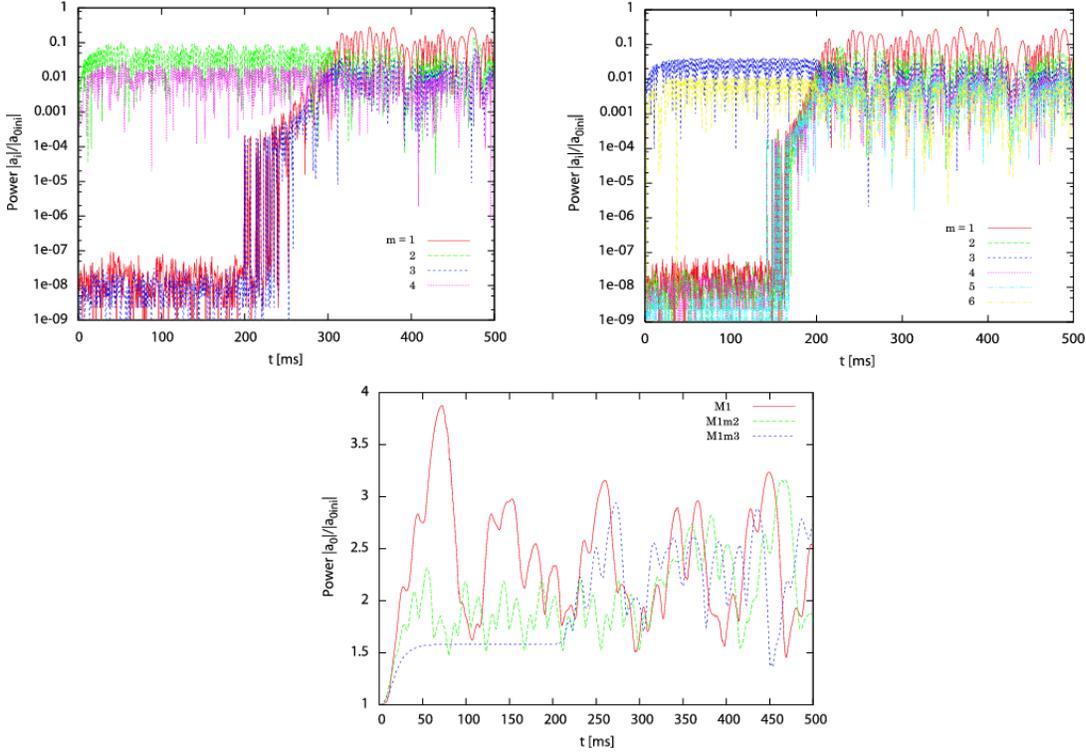}

\caption{The time evolutions of the amplitudes of various modes.
 The upper left (right) panel shows the results for Model M1m2 (M1m3).
 The lower panel displays the evolution of $m=0$ mode for Models
 M1, M1m2 and M1m3.}
\label{figmodedepegraph}
\end{figure}

 Next we show what happens if we impose initially the $m=2$ or $3$ mode
 instead of the $m=1$ mode, comparing Models M1, M1m2 and M1m3.
 In the upper panels of Figure~\ref{figmodedepegraph},
 we show the time evolutions of the amplitudes for various modes
 in Models M1m2 (left panel) and M1m3 (right panel).

 We first pay attention to the evolution up to $\sim 150$~ms,
 where the difference is most evident.
 Note that the linear phase lasts only for $\sim 10$~ms
 and the non-linear phase thereafter is the focus here.
 In the left panel, we see the growth and saturation of $m=4$ mode
 in addition to the original $m=2$ mode.
 On the other hand, the $m=6$ mode is formed to grow to the saturation
 in the right panel.
 Note that the $m=0$ mode is also generated in these models
 and will be discussed later.
 These models are produced by the non-linear mode-couplings,
 which are of quadratic nature,
 and the other modes with odd $m$ for Model M1m2,
 for example, are not generated.

 After $\sim 150$~ms, however, other modes also emerge and
 grow to be saturated.
 These modes are probably generated by numerical noises,
 having much smaller amplitudes initially and
 spending longer time in the linear phase.
 After the saturation of the modes,
 the dynamics is almost identical for the three models.

 In the lower panel of Figure~\ref{figmodedepegraph},
 we show the time evolution of the amplitudes of $m=0$ mode
 for the three models.
 As mentioned above,
 the modes are produced by the quadratic mode couplings
 of the initially imposed modes.
 Up to $\sim 150$~ms,
 the evolutions are quite different among them.
 For Model M1,
 the amplitude grows to be 4 times larger than
 the initial value in $\sim 50$~ms
 and oscillates violently thereafter.
 On the other hand,
 the maximum amplitudes are much smaller for Models M1m2 and M1m3,
 and the following oscillations have much smaller amplitudes.
 In fact, for Model M1m3 the amplitude becomes
 almost constant after $\sim 50$~ms.
 The saturation level is much lower than that of Model M1m2.
 It is interesting to point out that
 even though the $m=3$ mode is the unstable in the linear phase
 (see Figure~\ref{figlinem123eigen}),
 the average shock radius, or the $m=0$ mode, is
 most strongly affected by the $m=1$ mode.
 After $\sim 150$~ms, all modes are saturated
 and the behavior of the $m=0$ mode becomes
 almost identical among the different models.

\subsection{Dependence on the adiabatic index}
 In this paper, we employ the simple $\Gamma$-law EOS and so far we have 
 discussed only the case with $\Gamma=4/3$. In reality the EOS will not be 
 so simple and the adiabatic index may not be constant. In order to infer 
 the differences that the EOS may make, we vary the adiabatic index in the 
 the $\Gamma$-law EOS and see the changes in this subsection. 

 It is the unperturbed accretion flows that are most affected by the change of 
 adiabatic index. It is found that as the adiabatic index becomes larger,
 both the specific angular momentum and Bernoulli constant that allow the 
 existence of a standing shock wave gets smaller. This is understood as follows.
 The structure of unperturbed accretion flows and hence the existence of shock 
 are determined by the balance between the attractive gravity and the repulsive   
 centrifugal force and pressure. As the EOS becomes harder, 
 the pressure gets larger and, as a result, the centrifugal force can be reduced.
 For the same specific angular momentum, on the other hand, the Bernoulli constant
 can be smaller for harder EOS's. Note that the Bernoulli constant is a measure
 of the matter temperature at infinity.

 The instability itself is also affected by the change of adiabatic index, since
 it depends on the structure of unperturbed accretion flows.
 The results are summarized in Tables~\ref{tab:dyreso1} and \ref{tabreso1}. Although 
 it is difficult to find a systematic trend here, the saturation amplitude of the 
 $m=0$ mode appears to be correlated with the Mach number: as the Mach number 
 becomes larger by the change of adiabatic index, the saturation level gets enhanced.
 It is more important to understand here, however, that the instability does not change 
 qualitatively in spite of the relatively large variation in the injection parameters 
 that allow the existence of the standing shock wave. More thorough investigations
 of the EOS dependence will be a future task.
 
\subsection{Instability Mechanisms}

 To gain some insight into the instability mechanism,
 we calculate some time scales as follows:
\begin{eqnarray}
 \tau_{d-c(c)}  \equiv  \int_{r_{cor}}^{r_{sh}}
           \left( \, \frac{1}{|v^r|} + \frac{1}{C_s^{+}} \, \right) \, dr ,
 \label{eq:cycle_dcc}\\
 \tau_{c-c(c)}  \equiv  \int_{r_{cor}}^{r_{sh}}
           \left( \, \frac{1}{C_s^{-}} + \frac{1}{C_s^{+}} \, \right) \, dr ,
 \label{eq:cycle_ccc}\\
 \tau_{d-c(s)}  \equiv \int_{r_{inso}}^{r_{sh}}
           \left( \, \frac{1}{|v^r|} + \frac{1}{C_s^{+}} \, \right)  \, dr ,
 \label{eq:cycle_dcs}\\
 \tau_{c-c(s)}  \equiv \int_{r_{inso}}^{r_{sh}}
           \left( \, \frac{1}{C_s^{-}} + \frac{1}{C_s^{+}} \, \right) \, dr ,
 \label{eq:cycle_ccs}
\end{eqnarray}
 where $r_{inso}$ is the radius of inner sonic point and $C_s^{\pm}$ are the outgoing (+)
 and ingoing (-) sound velocities,
 respectively,
 and are given in the observer's frame for the Schwarzschild geometry as,
\begin{eqnarray}
 C_s^{\pm} = \Biggl| \, \frac{ \left(1-b_s^{\, 2}\right) u^t u^r \pm
 \biggl\{  \large\{ \left(1-b_s^{\, 2}\right) u^t u^r \large\}^2
    -[  \left(1-b_s^{\, 2}\right) \,\left( u^t \right)^2  \, -  \, b_s^{\, 2}  \, g^{tt} ] 
     [ \left(1-b_s^{\, 2}\right) \,\left( u^r \right)^2  \, -  \, b_s^{\, 2}  \, g^{rr} ] \biggr\}^{\frac{1}{2}}  }
          { \left(1-b_s^{\, 2}\right) \,\left( u^t \right)^2  \, -  \, b_s^{\, 2}  \, g^{tt} }
  \, \Biggr| . \label{eq:charaSch}
\end{eqnarray}
 Here $b_s$ denotes the sound velocity in the comoving frame.
 The corotation point of the perturbation is defined as
\begin{eqnarray}
 \omega_{r} - m \frac{u^{\phi}\left(r_{cor}\right)}{u^t\left(r_{cor}\right)} = 0 \label{eq:corodef}
\end{eqnarray}
 and its radius is expressed as $r_{cor}$.
 This investigation is inspired by the previous works \citep{fog2003,Gu2005}
 mentioned in the introduction.
 These time scales and the radius of corotation point for all the models
 are summarized in Table~\ref{tabreso1}.
 For comparison,
 we list the oscillation and growth time scales,
 which are obtained by linear analysis for
 the most unstable mode.

\begin{figure}
\includegraphics[angle=0,scale=.40]{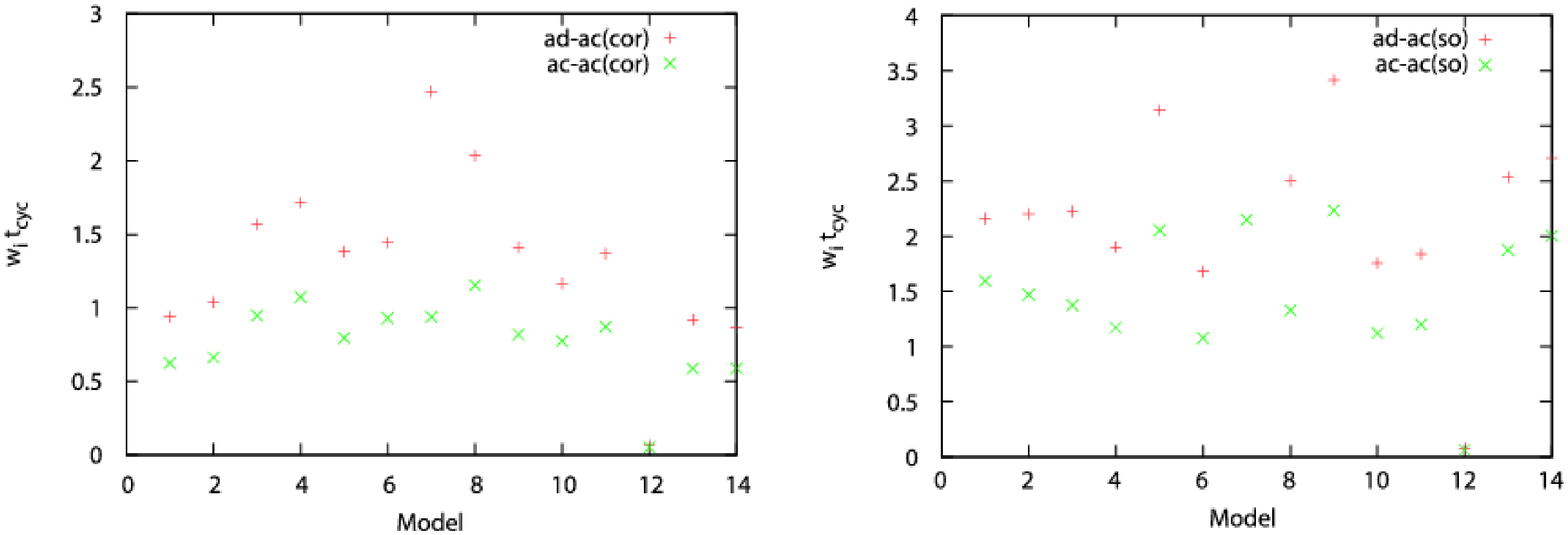}

\caption{The ratio of the growth rate to the frequencies of advective-acoustic (+)
 and acoustic-acoustic cycles ($\times$) for all the models.
 In the left (right) panel,
 the corotation (inner sonic) point is assumed to be the inner reflection point.}
\label{cyclefreq}
\end{figure}

 Figure~\ref{cyclefreq} compares
 the growth times with the cycle periods given
 by Eqs.~(\ref{eq:cycle_dcc})~to~(\ref{eq:cycle_ccs}) for all the models.
 It is found that the periods of acoustic-acoustic cycle
 are closer to the growth times than those of the
 advective-acoustic cycle,
 which appears to support the claim by
 the preceding papers \citep{fog2003,Gu2005}
 that the instability is of the Papaloizou-Pringle type.
 It is important, however, to point out that
 we can not identify the reflection point clearly
 (see Figure~\ref{cyclefreq}, in which
 the left (right) panel adopts the corotation (inner sonic) point
 as the inner reflection point).
 This is mainly because the wavelengths of the perturbations are rather long.
 In fact, we estimate the wavelength of acoustic perturbation by
\begin{eqnarray}
 \lambda_{w} \equiv \min \left( \frac{ 2 \pi b_s }{ \omega_{r} } \right) ,
 \label{eq:wlac}
\end{eqnarray}
 which should be much smaller than
 the scale height of the unperturbed flows
 for the justification of WKB approximations.
 The wavelength of the dominant unstable mode for each model is
 given in Table~\ref{tabreso1}.
 According to this estimation,
 it is comparable or longer than the scale height.
 Hence the WKB approximation is not justified at least
 for these models. Indeed, the reflection point of waves lose its meaning.
 Incidentally, the WKB approximation may be applicable to higher harmonics.
 In fact, there are sequences of unstable modes up to m=12 for Model M1, for example,
 and the wavelengths of their high harmonics are found to be shorter than the scale height. 
 It should be noted, however, that they have smaller growth rates and subdominant in
 driving the instability. Note also that the above analysis
 neither approves nor disproves of a particular
 mechanism in a mathematically rigorous sense.
 We need further investigation definitely.

\section{Implications for Astrophysical Phenomena}
 In the previous sections,
 we have found that the standing shock wave
 in the accretion flow into the Schwarzschild black hole
 is generally unstable to the non-axisymmetric perturbations
 and that it oscillates with large amplitudes
 in the non-linear regime.
 Here we consider the astrophysical implications of the shock instability,
 picking up Black Hole QPOs and GRBs as examples.
 
\subsection{Black Hole QPOs}
 As mentioned already,
 quasi-periodic oscillations have been observed
 for a couple of black hole candidates
 and they are attributed to some activities of
 the accretion disk around the black hole.
 The shock oscillation model for black hole QPOs
 has been investigated by many authors
 \citep{das03a,das03b,cha04,aoki04,oku07}.
 Recently, for example, \citet{oku07} performed
 two-dimensional pseudo-Newtonian numerical
 simulations of the shock oscillation
 in the meridian section assuming axisymmetry
 and taking into account the cooling and heating of gas
 and the radiation transport.
 They demonstrated that a quasi-periodically oscillating shock wave
 is formed around a black hole.
 They compared the numerical results with the observations
 for {\it GRS 1915+105} and suggested that
 the intermediate frequency QPO of this source
 might be due to the shock instability \citep{oku07}.

 Our models are different from those of \citet{oku07}.
 They consider the axisymmetric oscillations whereas
 we investigate the non-axisymmetric oscillations.
 We take general relativity fully into account.
 Besides, we calculate the energy density spectra for the present models.
 In so doing, we employ the data in the non-linear regime,
 that is, 100~ms after the onset of computations.
 Note that the dynamics in the non-linear phase is
 almost identical in all the models,
 including the ones, in which the $m=2$~or~$3$ mode
 is initially imposed instead of the $m=1$ mode.

\begin{figure}
\includegraphics[angle=0,scale=.60]{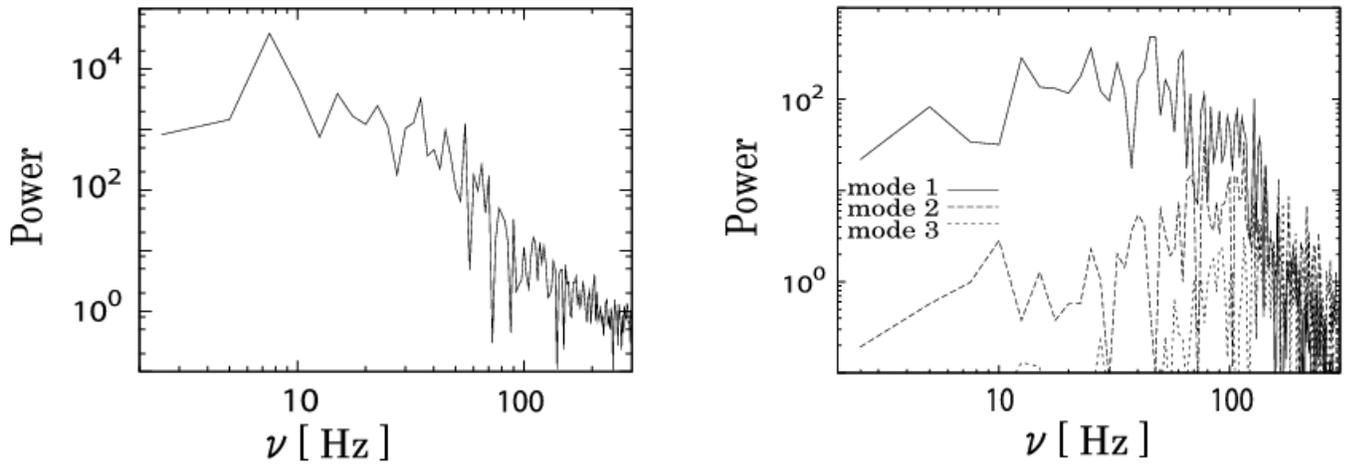}

\caption{Power spectra of energy density for the $m=0$ mode (left) and
 $m=1,2,3$ modes (right).}
\label{modefourierMode1}
\end{figure}

 Figure~\ref{modefourierMode1} shows
 the power spectra for the $m=0,1,2,3$ modes in Model M1.
 It is found that the $m=0$ mode has
 a quasi-periodic feature around 8~Hz,
 which corresponds to the period of
 large oscillations observed for the $m=0$ mode.
 Although there are some hints of other QPOs, they are 
 much less remarkable. This axisymmetric quasi-periodic oscillation
 is similar to those found by \citet{oku07}.
 The most important point here, however, is the fact that
 the quasi-periodicity of $m=0$ mode is
 induced by the non-axisymmetric instability through
 the quadratic mode coupling.
 
 Similar quasi-periodic oscillations are found also in other models.
 Their frequencies depend on the unperturbed flow
 and, hence, on the Bernoulli constant
 and specific angular momentum.
 The quantitative comparison with observations
 is beyond the scope of this paper,
 since we have neglected radiative processes, viscosity and,
 among other things,
 we have considered only the equatorial plane.
 It can be mentioned, however,
 that the non-axisymmetric shock instability is
 a good candidate of the source of QPOs
 and further investigations are certainly needed.

\subsection{Fluctuations in GRB jets}
 Long GRBs are currently thought to be 
 associated with massive stellar collapses
 and the subsequent formation of black holes.
 Although the central engine remains a mystery,
 it is widely believed that
 a highly relativistic jet is somehow produced near the black hole
 and its kinetic energy is later dissipated in internal shocks at larger distances,
 emitting gamma rays (see \citet{mes2006} for a recent review).
 In the so-called 'patchy shell' model,
 the jet actually consists of mass shells that have slightly different velocities
 and collide with each other, generating the internal shock waves.
 Although the time scale of the velocity fluctuations is thought
 to be set by the dynamical time scale of the black hole,
 the exact physical processes producing the velocity variations
 are unknown at present.

 During the collapse of massive stars giving rise to GRBs,
 a large amount of matter will accrete on a time scale of
 seconds onto a proto neutron star at first and into a black hole
 later. If the progenitor is rotating rapidly prior to the collapse
 \citep{yoon08},
 the accreting matter will form a disk around the compact object at the center.
 The accretion disk is expected to be
 advection-dominated
 \citep{pop98}.
 We are thus interested in the stability of the accretion flows
 into the black hole, especially in the accretion-dominated regime.
 Here we consider the accretion flows with a standing shock wave in them,
 since the core bounce produces a shock wave,
 which becomes an standing shock in the core soon after
 and will continue to exist in the subsequent accretions onto a proto neutron star,
 the phase just preceding the black hole formation. Even if the bounce shock does not
 survive, there will be a lot of chances of shock formation as long as 
 the standing shock is robust, since the velocity and pressure of accreting matter 
 are fluctuating in reality.

 According to the patchy shell model, gamma rays are emitted
 when the kinetic energy of ultra-relativistic jet
 is dissipated in internal shock waves,
 which are originated from the inhomogeneity of the jet.
 Although the mechanism of jet formation remains unknown,
 the black hole is supposed to be involved.
 The source of the inhomogeneity is also an unsolved problem.
 If a standing shock wave exists in the accretion flow,
 for example, as a relic of the shock wave produced at the core bounce,
 we speculate that the intrinsic instability of the system
 against non-radial perturbations
 will be a natural source of fluctuations in the GRBs jet
 if it is formed from some interactions between the accretion disk and black hole,
 which is not unlikely \citep{bla77}.
 It is mentioned incidentally that the recent progenitor models,
 which could produce GRB \citep{Mac99,her05},
 predict the injection parameters that are appropriate
 for the existence of a standing accretion shock.
 For example, \citet{her05} calculated the evolutions of massive stars, 
 taking into account magnetic fields and obtained 
 the specific angular momentum of several $\times 10^{16}$ cm$^2$/s and 
 the temperature $\lesssim 10^{10}K$ for the matter
 that will later form an accretion disk. These numbers are just 
 suitable for the existence of a standing shock wave in the accretion disk
 around a black hole of several M$_{\odot}$.

 Owing to the non-axisymmetric instability,
 the mass flux fluctuates very much indeed
 in our models.
 In this context, it is interesting to mention that
 the quasi-periodic oscillation with a period much longer than
 the dynamical time scale,
 which we have found in the previous sections, may leave
 its imprint somehow in the prompt gamma ray emissions or early X-ray afterglows. 
 It is certainly necessary, however,
 to study possible effects of cooling~\citep{pop98} on the instability.
 The disk thickness we ignored in this paper is also a concern in the future work.
 We finally mention that the gravitational radiation by the non-radial shock instability
 may also have interesting implications.

\section{Summary and Conclusion}
 We have investigated the non-axisymmetric shock instability
 in the accretion disk around Schwarzschild black holes,
 employing the fully general relativistic hydrodynamic simulations
 as well as the linear analysis.
 Both the linear and non-linear phases have been analyzed in detail.
 We have also given some possible implications
 for astrophysically interesting phenomena
 such as Black Hole QPOs and GRBs.

 The main findings in the present work are as follows:

 (a) The standing shock is generally unstable against non-axisymmetric
 perturbations, and a spiral arm structure is formed as a result of
 the growth of instability.
 It is typically one-armed, implying that the dominant mode in
 the non-linear phase is the $m=1$ mode.

 (b) In the linear phase,
 the dynamical simulations are in good agreement with
 the linear analysis in such features as stability, oscillation
 and growth time scales.
 The progressive modes,
 in which the deformed shock pattern rotates in the same direction
 as the unperturbed flow,
 are unstable and the retrogressive modes are stable.
 This is consistent with the previous works.

 (c) In the non-linear phase,
 various modes are produced by non-linear couplings,
 which are mainly of quadratic nature,
 and the amplitudes are saturated.
 The axisymmetric mode is
 also induced by the non-axisymmetric instability,
 and the shock radius oscillates with large amplitudes.
 The oscillation periods become slightly longer than
 in the linear analysis because of larger shock radii.

 (d) Even though strong perturbations are added initially,
 the shock remains to exist.
 Hence the disk plus shock system
 is quite robust in this sense.

 (e) The comparison of various cycle time scales
 with the linear growth times seems to support the claim
 that the instability is induced by the acoustic-acoustic cycle,
 although the inner reflection point is not identified unambiguously.
 It is important to note in this respect that
 the wavelength of perturbations is longer
 than the scale height,
 which does not allow the WKB approximation.

 (f) The Black Hole SASI found by \citet{molteni99}
 may be a promising candidate
 for the sources of the Black Hole QPOs and
 fluctuations in GRB jets.

 In the present study, we have also found that
 the non-axisymmetric instability is sensitive
 to the structure of the unperturbed steady flow.
 The general relativity is important in this respect.
 It should be stressed that
 the injection parameters that allow the existence
 of a standing shock wave
 are different between the GR
 and pseudo-Newtonian treatments. In fact, we have found by the direct comparison
 that the maximum specific angular momentum for the existence of multiple sonic 
 points is different by more than 60\% for the Bernoulli constant $E \le 1.003$. 

 Note also that the general relativity is indispensable in discussing
 the accretion into a Kerr black hole,
 since the frame-dragging will play an important role.
 This is currently undertaken \citep{nak08v2}.
 For more detailed comparison with observations,
 it is necessary to include the cooling and heating for GRBs case,
 and the magnetic field and viscosity for Black Hole QPOs.
 Last but not least, the discussed simulations including
 the polar dimension are inevitable.

\acknowledgments
We are grateful to Kenta Kiuchi and Yu Yamamoto for useful discussions. 
This work was partially supported by the Grant-in-Aid for the 21st century
COE program "Holistic Research and Education Center for Physics of
Self-organizing Systems" of Waseda University and for Scientific Research
of the Ministry of Education, 
Science, Sports and Culture of Japan (17540267, 14079202).

\appendix
\section{General Relativistic Hydrodynamic Code}
 Here we describe the GRHD code that are used in this paper.
 As mentioned already, it is base on the so-called central scheme,
 which guarantees a good accuracy
 even if flows include strong shocks and/or
 high Lorentz factors.
 Magnetic fields can be also included
 \citep{Del2002,shibata2005,duez2005}.

 Though we do not take into account the evolution of gravitational field,
 the so-called 3+1 formalism is suitable for hydrodynamics as well.
 Following~\citet{duez2005}, we write the metric in the form
\begin{eqnarray}
 ds^2 = - {\alpha}^2 dt^2 + \gamma_{ij}\left( dx^i + {\beta}^i dt \right)
 \left( dx^j + {\beta}^j dt \right) ,
\end{eqnarray}
 where $\alpha$, $\beta^i$, and $\gamma_{ij}$
 are the lapse, shift vector, and spatial metric, respectively.
 The basic equations for fluid dynamics in 3+1 form are expressed as:
\begin{eqnarray}
 \partial_t {\rho}_{\ast} + \partial_j\left( {\rho}_{\ast} v^j \right) & = & 0 ,
 \label{eq:conti3pra1} \\
 \partial_t S_i + \partial_j\left( \alpha \sqrt{\gamma} \, T^j \! _{i}\right) 
 & = & \frac{1}{2} \alpha \sqrt{\gamma} \, T^{\alpha \beta} g_{\alpha \beta ,i} ,
 \label{eq:Mon3pra1} \\
 \partial_t {\tau}
 + \partial_i \left( {\alpha}^2 \sqrt{\gamma} \, T^{0i} 
                - {\rho}_{\ast} v^i \right) & = & s ,
\end{eqnarray}
 where various variables are defined as follows:
\begin{eqnarray}
 v^j & \equiv & \frac{u^j}{u^t} , \label{eq:threevelodef} \\
 {\rho}_{\ast} & \equiv & \alpha \sqrt{\gamma} \, \rho_0 u^t ,
 \label{eq:con1def} \\
 S_j & \equiv & \alpha  \sqrt{\gamma} \, T^0 \! _j 
 = {\rho}_{\ast} h u_j ,
 \label{eq:con2def} \\
 \tau & \equiv & \alpha^2  \sqrt{\gamma} \, T^{00} - {\rho}_{\ast}
 = {\rho}_{\ast} \alpha h u^t -  \sqrt{\gamma} \, p - {\rho}_{\ast} ,
 \label{eq:con3def} \\
 s & \equiv & \alpha \sqrt{\gamma} \,
 \biggl\{ \left( T^{00} \beta^i \beta^j + 2 T^{0i} \beta^j + T^{ij} \right) K_{ij}
    -\left( T^{00} \beta^i + T^{0i} \right) \partial_{i} \alpha \biggr\} . \label{eq:defsourceMom}
\end{eqnarray}
 In the above equations,
 $\gamma$ and $K_{ij}$ are the determinant of the three metric
 and extrinsic curvature, respectively.
 We refer to $\rho_{\ast}$, $S_j$ and $\tau$
 as ``conserved variables (collectively denoted by $U$)'',
 whereas $\rho_0$, $p$ and $v^i$ are called
 ``primitive variables (collectively expressed as $P$)''.

 The conserved variables can be calculated
 directly from the primitive variables
 via Eqs.~(\ref{eq:con1def}),~(\ref{eq:con2def})~and~(\ref{eq:con3def}).
 There is no analytical expression for the primitive variables
 as a function of the conserved variables,
 on the other hand.
 Since we update the conserved variables
 rather than the primitive variables,
 we must need to solve the latter numerically at each time step
 because they are necessary for the calculations
 of the characteristic wave speed
 at each cell interface as shown later.
 If we use a $\Gamma$-law EOS,
 the inversion can be conducted easily
 as done by~\citet{duez2005}.
 The same method can not be applicable
 to the general EOS, however.
 Hence we take a different procedure based on the
 Newton-Raphson method, which will be explained below.

 We first write down a useful relation between $u^t$ and $u_j$
\begin{eqnarray}
 u^t = \frac{1}{\alpha} \biggl\{ 1 + \gamma^{ij} u_i u_j \biggr\}^{\frac{1}{2}} .
 \label{eq:utujrela}
\end{eqnarray}
 We define two more quantities as
\begin{eqnarray}
 f_1 & \equiv &  {\rho_0}^2 \gamma \biggl\{ {{\rho}_{\ast}}^2 h^2 + \gamma^{ij}  S_i S_j \biggr\}
 - {{\rho}_{\ast}}^4 h^2 , \label{eq:deff1} \\
 f_2 & \equiv &  \tau + {\rho}_{\ast} - {\rho}_{\ast} \alpha h u^t 
 +  \sqrt{\gamma} \,  p . \label{eq:deff2}
\end{eqnarray}
 We then search iteratively for the primitive variables that satisfy
 $f_1 = f_2 = 0$.
 We first guess two thermodynamical quantities $\rho_0$ and $p$.
 Then other thermodynamical quantities
 can be obtained from the EOS.
 Next, we obtain $u_j$ from Eq.~(\ref{eq:con2def})
 using $S_j$, $\rho_{\ast}$ and $h$,
 $u^t$ is determined by $u_j$ from Eq.~(\ref{eq:utujrela}).
 Thus the right hand sides of
 Eqs.~(\ref{eq:deff1})~and~(\ref{eq:deff2})
 are expressed as a function of
 only two thermodynamical quantities.
 We solve them by the Newton-Raphson method.
 The initial guess is obtained from the values at the previous step.

 The net flux at the cell interface is given
 by the approximate solution to the Riemann problem.
 Our code adopts the HLL (Harten. Lax, and van Leer) flux,
 which does not require the complete knowledge of the
 solutions to the Riemann problem but
 the maximum wave speed in each direction is needed.
 The first step for calculating the flux is to obtain $P_R$ and $P_L$,
 which are the values of primitive variables
 interpolated to the right- and left-hand side of
 each cell interface.
 We have implemented the MUSCL method \citep{HIRS1990} for this purpose.
 From $P_R$ and $P_L$,
 the maximum wave speed on each side of the cell interface,
 $c_{\pm,R}$ and $c_{\pm,L}$, can be calculated as in~\citet{duez2005}.

 The HLL flux is then expressed with the maximum wave speeds
 defined by $c_{+max} \equiv max (0,c_{+,R},c_{+,L})$
 and $c_{-max} \equiv max(0,c_{-,R},c_{-,L})$ as
\begin{eqnarray}
 f_{int} = \frac{ c_{-max} f_{R} +  c_{+max} f_{L} - c_{-max} c_{+max} \left( U_R - U_L\right) }
{ c_{-max} + c_{+max} } , \label{eq:HLLdef}
\end{eqnarray}
 where $f_R$ and $f_L$ are the fluxes
 calculated with $P_R$ and $P_L$, respectively.
 Note that if we define $c_{-max} = c_{+max} = max(0,c_{+,R},c_{+,L},c_{-,R},c_{-,L})$,
 then $f_{int}$ becomes the local Lax-Friedrichs flux.

\clearpage

\begin{deluxetable}{lcccccccc}
\tabletypesize{\scriptsize}
\rotate
\tablecaption{Model Parameters\label{tab1}}
\tablewidth{0pt}
\startdata
\hline\hline
   & Adiabatic & Bernoulli & Specific Angular & Inner Sonic & Shock Point & Mach Number & Initial
 & Initial Perturbation  \\
Model & Index \, $\Gamma$ & Constant \, $E$  & Momentum $\lambda$ & Point $r_{inso}$
 & $r_{sh}$ &  & Perturbation Mode & Amplitude \\
\hline
M1 & $\frac{4}{3}$ & 1.004 & 3.43$M_{\ast}$ &  5.3$M_{\ast}$ & 16.1$M_{\ast}$ & 2.4 & 1 & 1 \% \\
M2 & $\frac{4}{3}$ & 1.004 & 3.46$M_{\ast}$ &  5.2$M_{\ast}$ & 23.2$M_{\ast}$ & 2.3 & 1 & 1 \% \\
M3 & $\frac{4}{3}$ & 1.004 & 3.50$M_{\ast}$ &  5.0$M_{\ast}$ & 34.8$M_{\ast}$ & 2.1 & 1 & 1 \% \\
M4 & $\frac{4}{3}$ & 1.004 & 3.56$M_{\ast}$ &  4.8$M_{\ast}$ & 78.4$M_{\ast}$ & 1.5 & 1 & 1 \% \\
M5 & $\frac{4}{3}$ & 1.001 & 3.50$M_{\ast}$ &  5.1$M_{\ast}$ & 16.9$M_{\ast}$ & 4.1 & 1 & 1 \% \\
M6 & $\frac{4}{3}$ & 1.005 & 3.50$M_{\ast}$ &  5.0$M_{\ast}$ & 50.2$M_{\ast}$ & 1.6 & 1 & 1 \% \\
M7 & 1.033 & 1.13 & 3.80$M_{\ast}$ &  4.4$M_{\ast}$ & 38.7$M_{\ast}$ & 2.2 & 1 & 1 \% \\
M8 & 1.167 & 1.02 & 3.70$M_{\ast}$ &  4.6$M_{\ast}$ & 64.2$M_{\ast}$ & 1.4 & 1 & 1 \% \\
M9 & 1.167 & 1.02 & 3.60$M_{\ast}$ &  5.0$M_{\ast}$ & 14.0$M_{\ast}$ & 2.7 & 1 & 1 \% \\
M10 & 1.167 & 1.03 & 3.60$M_{\ast}$ &  5.0$M_{\ast}$ & 32.4$M_{\ast}$ & 1.5 & 1 & 1 \% \\
M11 & 1.433 & 1.001 & 3.35$M_{\ast}$ &  5.2$M_{\ast}$ & 40.6$M_{\ast}$ & 2.3 & 1 & 1 \% \\
M12 & 1.433 & 1.004 & 3.15$M_{\ast}$ &  6.0$M_{\ast}$ & 36.5$M_{\ast}$ & 1.3 & 1 & 1 \% \\
 \hline
M1m2 & $\frac{4}{3}$ & 1.004 & 3.43$M_{\ast}$ &  5.3$M_{\ast}$ & 16.1$M_{\ast}$ & 2.4 & 2 & 1 \% \\
M1m3 & $\frac{4}{3}$ & 1.004 & 3.43$M_{\ast}$ &  5.3$M_{\ast}$ & 16.1$M_{\ast}$ & 2.4 & 3 & 1 \% \\
 \hline
M1a10 & $\frac{4}{3}$ & 1.004 & 3.43$M_{\ast}$ &  5.3$M_{\ast}$ & 16.1$M_{\ast}$ & 2.4 & 1 & 10 \% \\
M1a100 & $\frac{4}{3}$ & 1.004 & 3.43$M_{\ast}$ &  5.3$M_{\ast}$ & 16.1$M_{\ast}$ & 2.4 & 1 & 100 \%
\enddata
\tablecomments{The locations of inner sonic point and shock surface are determined by
 the adiabatic index, Bernoulli constant and specific angular momentum.
 The mach number is calculated in the corotating observer's frame.
 $M_{\ast}$ is the mass of the central black hole.}
\end{deluxetable}

\begin{deluxetable}{lcccc}
\tabletypesize{\scriptsize}
\rotate
\tablecaption{Properties of Instability \label{tab:dyreso1}} 
\tablewidth{0pt}
\startdata
\hline\hline
   &  dominant mode    &  maximum amplitude   &  & \\
 Model  & in non-linear phase & of $m=0$ mode 
 &  $\tau_{la}$ & $\tau_{sm}$ \\
\hline
M1 & m=1 & 3.9 & $\approx$100ms & $\approx$20ms \\
M2 & m=1 & 5.1 & $\approx$120ms & $\approx$20ms\\
M3 & m=1 & 3.2 & $\approx$200ms & $\approx$20ms\\
M4 & m=1\hspace{0.5mm}or\hspace{0.5mm}2 & 1.3 & - & $\approx$ 50ms \\
M5 & m=1 & - &  - & - \\
M6 & m=1 & 1.5 & $\approx$210ms & $\approx$30ms \\
M7 & m=1 & 4.5 & - & $\approx$50ms \\
M8 & m=1 & 1.4 & $\approx$300ms & $\approx$20ms \\
M9 & m=1 & 3.5 & $\approx$80ms  $\approx$20ms \\
M10 & m=1 & 1.3 & - & $\approx$10ms \\
M11 & m=1 & 4.2 & $\approx$300ms & $\approx$60ms \\
M12 & - & - &  - & - \\
 \hline
M1m2 & m=1 & 3.1 & $\approx$100ms & $\approx$20ms \\
M1m3 & m=1 & 2.9 & $\approx$100ms & $\approx$20ms  \\
 \hline
M1a10 & m=1 & 3.5 & $\approx$100ms & $\approx$20ms \\
M1a100 & m=1 & 4.0 & $\approx$100ms & $\approx$20ms \\
\enddata
\tablecomments{$\tau_{la}$ ($\tau_{sm}$)
 is the large- (small-) amplitude oscillation period.
 The symbol (-) implies no identifications.}
\end{deluxetable}


\begin{deluxetable}{lcccccccc}
\tabletypesize{\scriptsize}
\rotate
\tablecaption{Cycle Frequencies \label{tabreso1}} 
\tablewidth{0pt}
\startdata
\hline\hline
   & Corotation Point & Oscillation Period & Growth Time & Wavelength of Acoustic Perturbations 
   &  & & & \\
Model & $r_{coro}$ & $t_{osci}$  & $t_{grow}/2\pi$ & $\lambda_{w}$ & $\tau_{d-c(c)}$ 
   & $\tau_{c-c(c)}$ & $\tau_{d-c(s)}$ & $\tau_{c-c(s)}$ \\
\hline
M1 & 47.7km (10.6$M_{\ast}$) & 3.7ms & 3.2ms & 142.6km (31.7$M_{\ast}$) & 3.0ms & 2.0ms & 6.9ms & 5.1ms \\
M2 & 64.3km (14.3$M_{\ast}$) & 6.4ms & 4.8ms & 213.7km (47.5$M_{\ast}$) & 5.0ms & 3.2ms & 10.6ms & 7.1ms \\
M3 & 62.3km (13.9$M_{\ast}$) & 5.9ms & 8.2ms & 171.9km (38.2$M_{\ast}$) & 12.9ms & 7.8ms & 18.3ms & 11.3ms \\
M4 & 61.2km (13.6$M_{\ast}$) & 5.6ms & 29.0ms & 120.1km (26.7$M_{\ast}$) & 49.7ms & 31.1ms & 55.1ms & 33.9ms \\
M5 & 50.4km (11.2$M_{\ast}$) & 4.1ms & 2.4ms & 147.6km (32.8$M_{\ast}$) & 3.3ms & 1.9ms & 7.5ms & 4.9ms \\
M6 & 53.1km (11.8$M_{\ast}$) & 4.5ms & 17.3ms & 114.3km (25.4$M_{\ast}$) & 25.0ms & 16.1ms & 29.0ms & 18.6ms \\
M7 & 104.8km (23.3$M_{\ast}$) & 14.5ms & 8.4ms & 294.3km (65.4$M_{\ast}$) & 20.8ms & 7.9ms & 159.2ms & 18.1 ms \\
M8 & 61.2km (13.6$M_{\ast}$) & 5.5ms & 22.6ms & 121.5km (27.0$M_{\ast}$) & 46.0ms & 26.1ms & 56.6ms & 30.0ms \\
M9 & 43.6km (9.7$M_{\ast}$) & 3.1ms & 2.2ms & 97.2km (21.6$M_{\ast}$) & 3.1ms & 1.8ms & 7.5ms & 4.9ms \\
M10 & 61.2km (13.6$M_{\ast}$) & 5.6ms & 11.0ms & 157.9km (35.1$M_{\ast}$) & 12.8ms & 8.5ms & 19.3ms & 12.3ms \\
M11 & 67.5km (15.0$M_{\ast}$) & 7.2ms & 10.7ms & 211.0km (46.9$M_{\ast}$) & 14.6ms & 9.3ms & 19.6ms & 12.8ms \\
M12 & 44.1km (9.8$M_{\ast}$) & 3.5ms & 252.2ms & 113.8km (25.3$M_{\ast}$) & 16.1ms & 13.0ms & 18.9ms & 15.6ms \\
 \hline
M1m2 & 51.7km (11.5$M_{\ast}$) & 2.2ms & 2.7ms & 82.8km (18.4$M_{\ast}$) & 2.5ms & 1.6ms & 6.9ms & 5.1ms \\
M1m3 & 54.0km (12.0$M_{\ast}$) & 1.5ms & 2.6ms & 58.9km (13.1$M_{\ast}$) & 2.2ms & 1.5ms & 6.9ms & 5.1ms \\
\enddata
\tablecomments{
 $t_{osci}$, $t_{grow}/2\pi$ and $\lambda_{w}$ represent
 the oscillation period, growth time and wavelength of acoustic perturbations, respectively,
 which are obtained by linear analysis.
 $\tau_{d-c(c)}$, $\tau_{c-c(c)}$, $\tau_{d-c(s)}$, $\tau_{c-c(s)}$ are obtained from
 Eqs. (\ref{eq:cycle_dcc})~to~(\ref{eq:cycle_ccs}), which show the acoustic-acoustic cycle
 or advective-acoustic cycle between the shock surface and the corotation or inner sonic point.
 }
\end{deluxetable}

\clearpage

\end{document}